\documentclass[12pt]{article} % use larger type; default would be 10pt
\usepackage[utf8]{inputenc} % set input encoding (not needed with XeLaTeX)

\usepackage{geometry} % to change the page dimensions
\usepackage{graphicx} % support the \includegraphics command and options
\usepackage{graphicx}
\usepackage{caption}
\usepackage{cite}
\usepackage{color} 
\usepackage{epstopdf}
\usepackage{rotating}
\usepackage{lscape}
\usepackage{lineno}
\usepackage{multirow}
\usepackage{amsmath} 
\usepackage[table]{xcolor}
\usepackage{breakcites}
\usepackage{amsfonts}
\usepackage{booktabs} % for much better looking tables
\usepackage{array} % for better arrays (eg matrices) in maths
\usepackage{paralist} % very flexible & customisable lists (eg. enumerate/itemize, etc.)
\usepackage{verbatim} % adds environment for commenting out blocks of text & for better verbatim
\usepackage{subfig}
\usepackage{bm}
\usepackage{fancyhdr} % This should be set AFTER setting up the page geometry

\usepackage[numbers]{natbib}

\setcitestyle{open={},close={}}

\usepackage{caption} 
\usepackage{setspace} 

\usepackage{authblk}
\usepackage{sectsty}
\allsectionsfont{\sffamily\mdseries\upshape}

\usepackage[nottoc,notlof,notlot]{tocbibind} % Put the bibliography in the ToC
\usepackage[titles,subfigure]{tocloft} % Alter the style of the Table of Contents

\title{\textbf{Escape cascades as a behavioral contagion process with adaptive network dynamics}} 

\date{}
      
\author[1,2]{Wenhan Wu}
\author[1,$*$]{Xiaoping Zheng}
\author[2,3,4,$*$]{Pawel Romanczuk}
\affil[1]{\textit{Department of Automation, Tsinghua University, Beijing, China.}}
\affil[2]{\textit{Institute for Theoretical Biology, Department of Biology, Humboldt Universität zu Berlin, Berlin, Germany.}}
\affil[3]{\textit{Research cluster of excellence "Science of Intelligence", Technische Universität Berlin, Berlin, Germany.}}
\affil[4]{\textit{Bernstein Center for Computational Neuroscience, Berlin, Germany.}}

\begin{document}
\maketitle    

\footnote{*asean@mail.tsinghua.edu.cn}
\footnote{*pawel.romanczuk@hu-berlin.de}

\newpage
\section*{Abstract}
Complex behavioral contagion in collective evasion of mobile animal groups can be predicted by reconstructing quantitative interaction networks. Based on the assumption of time-scale separation between a fast contagion process and a slower movement response, the underlying interaction networks have been previously assumed to be static, determined by the spatial structure at the onset of the collective escape response. This idealization does not account for the temporal evolution of the spatial network structure, which may have a major impact on the behavioral contagion dynamics. Here, we propose a spatially-explicit, agent-based model for the coupling between behavioral contagion and the network dynamics originating from the spreading movement response. We explore the impact of movement parameters (startle speed, initial directionality, and directional noise) on average cascade size. By conducting numerical simulations for different density levels, we show that increasing escape speed suppresses the cascade size in most cases, that the cascade size depends strongly on the movement direction of the initially startled individual, and that large variability in the direction of individual escape movements (rotational noise) will typically promote the spread of behavioral contagion through spatial groups. Our work highlights the importance of accounting for movement dynamics in behavioral contagion, and facilitates our understanding of rapid coordinated response and collective information processing in animal groups.

\newpage
\section*{Introduction}  
Collective behavior can be observed in a wide range of biological and social systems \cite{Vicsek2012,BakColeman2021}, ranging from animal groups (e.g., bird flocks \cite{Cavagna2010}, fish schools \cite{GomezNava2023}, sheep herds \cite{Ginelli2015}, and insect swarms \cite{Sarfati2021}) to human societies (e.g., crowd evacuation \cite{Wu2022}, group decision making \cite{Tang2021}, and social activity patterns \cite{Barabasi2005,Schlaepfer2021}). In general, complex collective behaviors emerge in a self-organized way from the interplay of various factors, such as individual-level perception and information processing, and the interactions between individuals within the collective \cite{Couzin2009,Sumpter2005}. Collective evasion maneuvers in mobile animals represent a special type of collective dynamical behavior, manifesting itself via the fast spread of individual escape reactions through the collective \cite{Rosenthal2015, Poel2021}. While being biologically highly relevant, this behavior is also comparably easy to quantify experimentally due to the typically clearly identifiable, stereotypical escape movements. The research related to such behavioral cascades, which can be considered a form of collective information processing, has been attracting considerable interest in recent years \cite{Kastberger2008,Procaccini2011,Rosenthal2015, Poel2021, GomezNava2023}. In particular, local and fast changes in the state of an initially startled individual (e.g., a reflexive behavior to quickly move away from danger in response to light, sound, or predator stimuli) give rise to similar behaviors in surrounding neighbors, and the escape waves caused by behavioral contagion can propagate across the group \cite{Rosenthal2015,Poel2022}. This instability towards sudden behavioral changes has been confirmed to be an inherent property of mobile groups, which can be beneficial for optimizing their collective access to environmental information and promoting the rapid transmission of directional information \cite{Buhl2006}.

To move beyond a descriptive understanding of the social transmission of behavioral contagion in collective evasion maneuvers, several studies have adopted computational models to reproduce experimental observations and predict behavioral cascades. The fundamental nature of social contagion in schooling fish was successfully revealed by reconstructing interaction networks of behavioral propagation based on the sensory information, making it possible to predict the magnitude of behavioral cascades across groups before they occur \cite{Rosenthal2015}. Further, a modeling approach based on a generalized (complex) contagion model \cite{Dodds2004,Dodds2005} was developed to demonstrate that average cascades strongly depend on changes in spatial positioning rather than changes in individual responsiveness \cite{Sosna2019}. Using a similar susceptible-infected-recovered (SIR) type model for behavioral contagion, it has been shown that groups can manage a trade-off between sensitivity and robustness in collective information processing according to the riskiness and noisiness of the environment \cite{Poel2022}. In the above studies the interaction network has been assumed static, determined by the spatial structure at the onset of the first escape response. This assumption can be justified by the time scale separation between the fast contagion process and the comparably slow individual movements. The assumption is reasonable in particular at the early stages of the spreading process, and thus predicts well smaller cascades with shorter duration as typically observed in laboratory experiments \cite{Poel2022}. However, the individual escape movements triggered by the spreading startle response may strongly alter the spatial network structure. This leads to changes in the visual information captured by neighboring individuals, which in turn feeds back into the further spreading of behavioral cascades \cite{Kulahci2018}. Therefore, the static network approximation becomes questionable in particular for larger cascades with longer duration, during which significant rearrangements of the spatial structure take place. Thus, in order to fully understand the collective escape response, it becomes important to explicitly account for the feedback between the contagion process and the dynamic network structure.  

Here, we investigate how movement responses spreading in groups via a contagion process feed back on the contagion dynamics itself using an individual-based model in two spatial dimensions. While our work has been directly inspired by escape cascades in fish schools, it is of general relevance for all systems where the contagion process triggers individual movement responses, which in turn affects the interaction network. Note, that a model recently proposed by Levis \emph{et al.} \cite{Levis2020} considers also the interplay of a susceptible-infected-susceptible (SIS) epidemic process and collective movement. However, besides considering a simpler contagion process, the work focuses on flocking dynamics, where agents are permanently moving and the infections spread a desired direction of motion, and do not induce the movement response itself, as considered here. 

Deviating from previously considered models of escape cascades \cite{Sosna2019,Poel2022}, we propose a coupling model that considers the behavioral cascades spreading on top of individual movements to fully capture key motion characteristics during the contagion process. Our simulation results indicate that the propagation range of behavioral cascades depends strongly on the specific properties of each movement parameter. It can be observed that a relatively slow constant speed, an initial movement direction towards the group's center of mass, and a higher intensity of the rotational noise will facilitate the spread of behavioral contagion through spatial groups. By comparing the contagion dynamics at different density levels, we find that there are also differences in the impact of different movement parameters on the cascade size depending on the group density. In summary, this work emphasizes that the complete spatio-temporal dynamics of interaction networks for behavioral contagion should be considered, for an encompassing understanding of information propagation in dynamical, spatially-embedded groups.

\section*{Results} 

\subsection*{Contagion model with movement response}

We consider a system of $N$ agents, e.g., individuals in fish school, which are represented by nodes in an interaction network. They can be in one of three states $S_i(t)$: susceptible, active, and recovered. 
The behavioral contagion model part corresponds to a continuous-time variant of the Dodds \& Watts model \cite{Dodds2004} introduced in \cite{Sosna2019}.
The process of behavioral contagion can be described as below: A susceptible individual $i$ receives stochastic doses of activation signal of magnitude $d_a$ from an active neighbor $j$ (note that neighbors are determined by visual networks \cite{Poel2021}) at a rate $r_{ij} = r_{\max}w_{ij}$, which is proportional to the link weight $w_{ij}$ in the interaction network, and $r_{\max}$ is the maximum rate of sending activation doses for $w_{ij} = 1$. For individual $i$, the stochastic time series of activation signal receiving from an active neighbor $j$ is given by:
\begin{equation}
d_{ij}(t) = \left\{ {\begin{array}{*{20}{c}}
{{d_a},{\kern 1pt}{\kern 1pt}{\kern 1pt}{\kern 1pt}{\kern 1pt}{\kern 1pt}{\kern 1pt}{\kern 1pt}{\kern 1pt}{\kern 1pt}{\kern 1pt}{\kern 1pt}{\kern 1pt}{\kern 1pt}{\kern 1pt}{\kern 1pt}{\kern 1pt}{\kern 1pt}{\kern 1pt}{\kern 1pt}{\kern 1pt}{\kern 1pt}{\kern 1pt} {p_a}}\\
{0,{\kern 1pt}{\kern 1pt}{\kern 1pt}{\kern 1pt}{\kern 1pt}{\kern 1pt}{\kern 1pt} 1 - {p_a}}
\end{array}} \right.
\end{equation}
where $p_a = r_{ij}\Delta t$ is the probability of receiving an activation dose within a short time step $\Delta t$. From this, the cumulative dose $D_i(t)$ is updated by integrating its recent memory in the form of exponential decay:
\begin{equation}
\frac{{d{D_i}(t)}} {dt} =  - \delta {D_i}(t) + \frac{1}{\Delta t}\sum\limits_j {d_{ij}(t)} 
\label{eq:dosedynamics}
\end{equation}
where $\delta $ is a discount factor. By using a standard Euler discretization, Equation \ref{eq:dosedynamics} can be rewritten as follows:
\begin{equation}
{D_i}(t) = (1 - \delta \Delta t){D_i}(t - \Delta t) + \sum\limits_j {d_{ij}(t)} 
\end{equation}
If the cumulative dose $D_i(t)$ exceeds the internal response threshold $\theta$ of individuals, it will enter the active state and start performing an escape movement with a constant speed $v_0$ in the direction $\vec e_i$, copying the movement direction of its active neighbors with directional noise:
\begin{equation}
{\vec e_i} = \frac{{\langle {\vec e_j} \rangle_j + {\vec \sigma}_i}} {\left\| {\langle {\vec e_j} \rangle_j + {\vec \sigma }_i} \right\|}
\end{equation}
Here, $\langle {\vec e_j} \rangle_j$ stands for the normalized average direction of active neighbors $j$, $\vec \sigma_i = {\sigma_0} \cdot (\cos\eta ,\sin\eta)$ represents the rotational noise, where ${\sigma_0}$ is the noise intensity, and $\eta$ is a random number taken from a uniform distribution $U(0,2\pi)$. After a fixed activation time $\tau_{act}$, individual $i$ will change its state from active to recovered and stop until the end of the simulation.

\subsection*{Construction of interaction networks} 
The model of the interaction network $A_{ij} = (w_{ij})_{n \times n}$, by which the escape behavior propagates across groups, has been derived from empirical data, based on the behavior of first responders after an initial, spontaneous startle \cite{Rosenthal2015,Sosna2019,Poel2022}. This has been achieved by analyzing a series of relative features of the initially startled individual from the perspective of the responder. It has been shown that in a logistic regression model the most predictive feature of behavioral response was the logarithm of the metric distance. The second most predictive feature was the ranked angular area of the initially startled individual occupied on the field of vision of the responding individual, with a much weaker contribution (smaller coefficient). 

Here, for simplicity, we calculate the link weights $w_{ij}$  of the interaction solemnly based on the logarithmic distance, while still accounting for visual occlusions by explicitly calculating visual networks \cite{Poel2021}. There is no link ($w_{ij}=0$) if a neighbor $j$ occupies a fraction of the visual field of the focal individual $i$ below the visual threshold $\theta_{vis}$. 

Following the logistic regression model, the link weight $w_{ij}$ can be calculated as the probability of first response by individual $i$ to a single initial startle by individual $j$, as given by the following equation:
\begin{equation}
w_{ij} = \frac{1} {1 + \exp[ - \beta_1 - \beta_2\log(l_{ij})]}
\end{equation}
Here, $\beta_1$ and $\beta_2$ are the fitting coefficients obtained previously by performing a logistic regression on experimental data \cite{Sosna2019}, and $l_{ij} = \| {\vec x}_i - {\vec x}_j \|$ is the Euclidean distance between individuals $i$ and $j$. Due to the movement response of activated individuals, their position vectors will evolve in time, and thus also the relative distances will be time-dependent: $l_{ij}=l_{ij}(t)$. Therefore, the corresponding networks are dynamic and continuously recalculated during the contagion process.

\subsection*{Model parameters} 
All model parameters are listed in Table 1. To be consistent with previous experimental work and corresponding models~\cite{Rosenthal2015,Sosna2019,Poel2022}, we fix the values of a range of parameters. The number of individuals $N = 40$ is set to the experimental value in \cite{Sosna2019}, and the visual threshold $\theta_{vis} = 0.02{\kern 1pt}rad$ is consistent with the previously chosen value. The coefficients $\beta_1 = -0.271$ and $\beta_2 = -2.737$, are taken from a logistic regression of first responders. The maximal rate $r_{\max} = 10^2{\kern 1pt}s^{-1}$, numerical time step $\Delta t = 0.01{\kern 1pt}s$, magnitude of activation dose ${d_a} = 10^{-2}$, and activation time $\tau_{act} = 1.0{\kern 1pt}s$ in the behavioral contagion model are referred to \cite{Sosna2019}. The discount factor $\delta = 0.1$ provides a reasonable decay rate for the cumulative dose. In addition to the parameters with fixed values, the following parameters have been varied throughout our simulations, the activation threshold $\theta_{act}$, by being set to a critical response threshold for each density level, constant speed $v_0$, initial movement direction $\alpha_0$ relative to the group's center of mass, and directional noise intensity $\sigma_0$. All distances are measured in terms of the body length of the agents (long axis if the ellipsoid body). The ellipsoid shape with a fixed aspect ratio of 0.4 is motivated by the the body form of fish, but the general result presented here do not depend on a specific choice of the aspect ratio.

\begin{table}
	\centering
	\captionsetup{justification=centering}  
	\caption{{Model parameters with values fixed throughout this study.}}
	\setlength{\tabcolsep}{12mm}{
	\begin{tabular}{ccc}      
		\hline 
		\hline
		\textbf{Symbol} & \textbf{Description} & \textbf{Value}\\ \hline
		$N$ & number of individuals & $40$ \\
  		$\theta_{vis}$ & visual threshold & $0.02{\kern 1pt}rad$ \\
		$\beta_1$ & intercept & $-0.271$ \\
		$\beta_2$ & LMD coefficient & $-2.737$ \\
		$r_{max}$ & maximal rate & $10^2{\kern 1pt}s^{-1}$ \\
		$\Delta t$ & numerical time step & $0.01{\kern 1pt}s$ \\
		$d_a$ & magnitude of activation dose & $10^{-2}$ \\
		$\tau_{act}$ & activation time & $1.0{\kern 1pt}s$ \\
		$\delta$ & discount factor & $0.1$ \\
        $\theta_{act}$ & activation threshold & varied \\
        $v_0$ & constant speed & varied \\
        $\alpha_0$ & initial movement direction & varied \\
        $\sigma_0$ & directional noise intensity & varied \\
		\hline 
		\hline
    \label{tab:parameters}
	\end{tabular}} 
\end{table}

\subsection*{Coupling of behavioral contagion and movement dynamics} 
\begin{figure}[tbp]
	\centering
	\includegraphics[width=10cm]{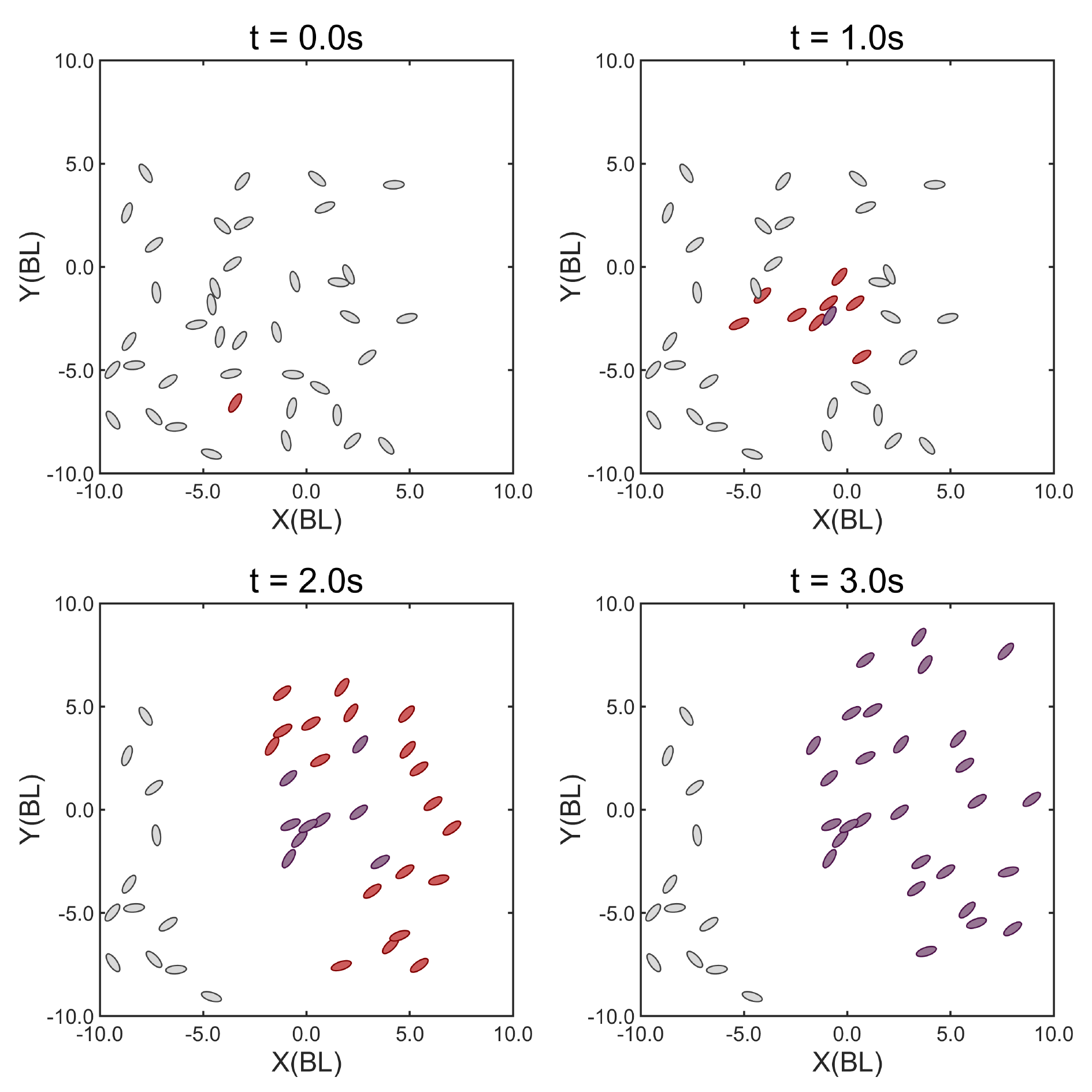}
	\begin{spacing}{1.0}
	\caption{Spatio-temporal snapshots of a single startle cascade generated by a group of 40 individuals. The susceptible, active, and recovered individuals are represented by gray, red, and purple ellipses, respectively.}
	\label{Fig.1}
	\end{spacing}
\end{figure}

Previous models with static interaction networks \cite{Rosenthal2015,Sosna2019,Poel2022} account well for the initial stages of escape cascades. However, the idealized assumption results in the actual movement of individuals, with its consequences for the interaction network, being entirely ignored. By explicitly taking into account the coupling between behavioral contagion and the corresponding movement response, we formulate a model of dynamically evolving, adaptive interaction networks \cite{Gross2009} governing the dynamics of collective escapes. Fig. \ref{Fig.1} shows spatio-temporal snapshots of a single startle cascade generated by a group of 40 individuals ($v_0 = 5{\kern 1pt}BL/s$, $\alpha_0 = 58.4^\circ$, $\sigma_0 = 0.3$). At the initial time ($t = 0.0{\kern 1pt}s$), one individual starts out as an active state (red), while the others are in a susceptible state (gray). As the initially startled individual moves towards the upper right, the link weights between the initially startled individual and a subset of its neighbors increase due to the decreasing distance, which leads to an increasing probability of receiving activation signals. These neighbors will become more likely active if the cumulative dose exceeds the response threshold (here we set $\theta  = 0.1$). When they initiate an escape themselves, their movement directions approximately match the one of the initially startled individual ($t = 1.0{\kern 1pt}s$). The collective evasion spreads further over time ($t = 2.0{\kern 1pt}s$), with the later activated individuals moving towards the upper right corner, while the early activated individuals enter the recovered state (purple) after a fixed activation time $\tau_{act} = 1.0{\kern 1pt}s$. Eventually, all activated individuals in the fish group become recovered and stop moving ($t = 3.0{\kern 1pt}s$).

\begin{figure}[tbp]
	\centering
	\includegraphics[width=12cm]{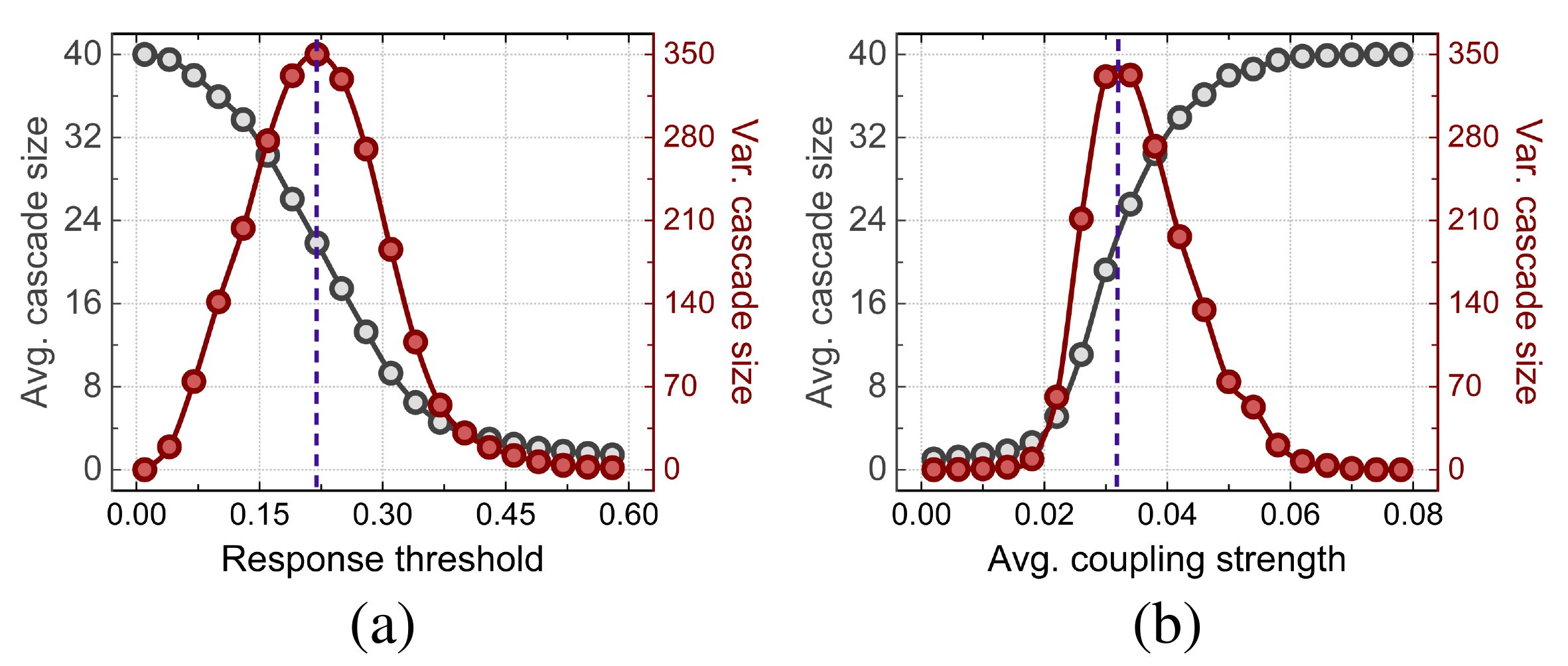}
	\begin{spacing}{1.0}
	\caption{Phase transition from local to global cascades triggered by two control parameters. (a) The average and the variance of cascade sizes as a function of the response threshold. (b) The average and the variance of cascade sizes as a function of the average coupling strength. The blue dashed line corresponds to the quasi-critical point of the phase transition.}
	\label{Fig.2}
	\end{spacing}
\end{figure}

An important question is what are the control parameters that are able to induce the phase transition from local to global cascades in such systems. In previous studies with static interaction networks, the response threshold $\theta$ (i.e., the responsiveness to social cues) and average coupling strength $\langle w_{ij} \rangle$ (i.e., the average link weights) are the two relevant control parameters. Fig. \ref{Fig.2}(a) shows the average and the variance of cascade sizes as a function of the response threshold. At low response thresholds, the cascade propagates rapidly across the majority of the group and we observe large cascades with almost all individuals participating. As the threshold increases, susceptible individuals need to receive more activation signals to be active, and the startles fail to propagate. The average cascade size decreases sharply, and eventually only small local cascades around the initially startled individual can be observed. The variance of the cascade size distribution has a maximum at the intermediate value of response thresholds. The corresponding response threshold is known as the quasi-critical point (blue dashed line) of phase transition in a finite-sized system. In Fig. \ref{Fig.2}(b), we show that the impact of increasing the average coupling strength on cascading behavior is directly opposite to that of the response threshold. At low coupling strength, the weak social signals are not sufficient for the cascade to propagate. With increasing coupling the network structures become more tightly connected, and the cumulative activation dose of a susceptible individual neighboring an active one grows faster, which results in an increasing frequency of large cascades. Again, the maximum value in the variance of cascade size indicates the critical coupling strength in a finite-sized system. Note that with increasing system size the location of the quasi-critical control parameters shifts, and the actual critical point in the thermodynamic limit has to be estimated from finite size scaling \cite{Attanasi2014}. 

The influence of the above two control parameters on collective evasion is apparent and rather straightforward to understand, however, new parameters related to movement dynamics may also have a non-negligible impact if we consider adaptive, spatially-embedded interaction networks \cite{Peruani2008,Zhao2022}. For instance, the constant speed $v_0$ of individuals can play a crucial role in behavioral contagion: If we define a contact of two individuals as instances where they are in close proximity, within a well-defined distance threshold, then a slow individual will be in contact with few neighbors within its activation time, but with each of them it will have long interaction times in comparison to a fast individual, which can potentially establish a larger number of rather short-lived contacts. Another potential parameter is the initial movement direction $\alpha_0$, because an active individual startling at the periphery of the group will establish stronger connections with more individuals if it moves towards the group's center of mass. Conversely, the connections with other individuals become weaker, if the focal individual moves out of the group (i.e., away from the group's center of mass), which in turn should have an inhibitory effect on behavioral cascades. In addition, we may also consider ``errors'' in copying the startle direction controlled by a noise intensity $\sigma_0$, which can also be an important factor. A higher noise will actually inhibit the alignment of startle movements of different individuals, but its impact on behavioral contagion may vary depending on movement characteristics. Therefore, we will investigate how the three parameters governing the movement response: startle speed, initial directionality, and directional noise affect collective evasion in the following subsection.

\begin{figure}[tbp]
	\centering
	\includegraphics[width=17cm]{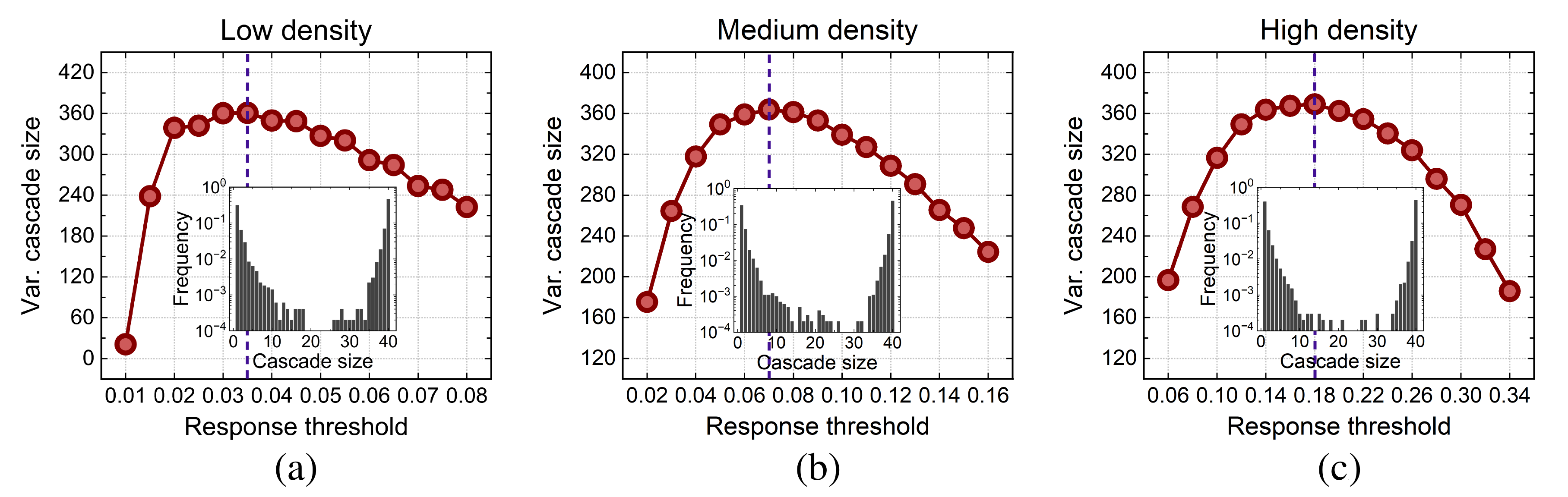}
	\begin{spacing}{1.0}
	\caption{Determination of activation thresholds (i.e., the critical response thresholds) at different density levels. (a) Low density. (b) Medium density. (c) High density. The inset shows the distribution of cascade sizes at the activation threshold.}
	\label{Fig.3}
	\end{spacing}
\end{figure}

\subsection*{Effects of movement parameters on behavioral cascades}
The average cascade size has been confirmed to be mainly modulated by changes in typical inter-individual distance \cite{Poel2022}. Therefore, we construct fish groups of $N=40$ individuals (as in experiments \cite{Sosna2019}), with three density levels, by initially distributing the individuals in rectangular areas of size $20\times20{\kern 1pt}BL^2$ (low density, $\rho=0.1{\kern 1pt}BL^{-2}$), $15\times15{\kern 1pt}BL^2$ (medium density, $\rho=0.18{\kern 1pt}BL^{-2}$), and $10\times10{\kern 1pt}BL^2$ (high density, $\rho=0.4{\kern 1pt}BL^{-2}$), respectively. To estimate the critical point of phase transition from local to global cascades at each density, we repeat the simulation 50 times for different networks at each candidate response threshold, which generates the distribution of cascade sizes. From this, we calculate the corresponding variance of cascade size, and the (quasi-)critical point in a finite-sized system corresponds to the value of the response threshold with a maximum variance. At the critical response threshold, variations of other control parameters will have the largest impact on the cascade sizes, thus these are the optimal thresholds for investigating the role of movement-related parameters on the behavior of the system. Fig. \ref{Fig.3} shows the determination of activation thresholds (i.e., the critical response thresholds) at different density levels (low density: $\theta_{act} = 0.035$, medium density: $\theta_{act} = 0.07$, high density: $\theta_{act} = 0.18$). A maximum of the cascade size variance can be clearly observed for all densities, and the corresponding activation threshold grows with the increase in density. Given that the average coupling strength of the group with a higher density is larger, it is easier to trigger global cascades under the same response threshold, hence the activation threshold needs to be increased to produce more local cascades. The distribution of cascade sizes at the activation threshold indicates that approximately half of the cascades remain relatively small, while the other half spreads across the majority of the group. It is notable that the bimodality of the cascade size distribution appears to become stronger with increasing density level.

\begin{figure}[tbp]
	\centering
	\includegraphics[width=15cm]{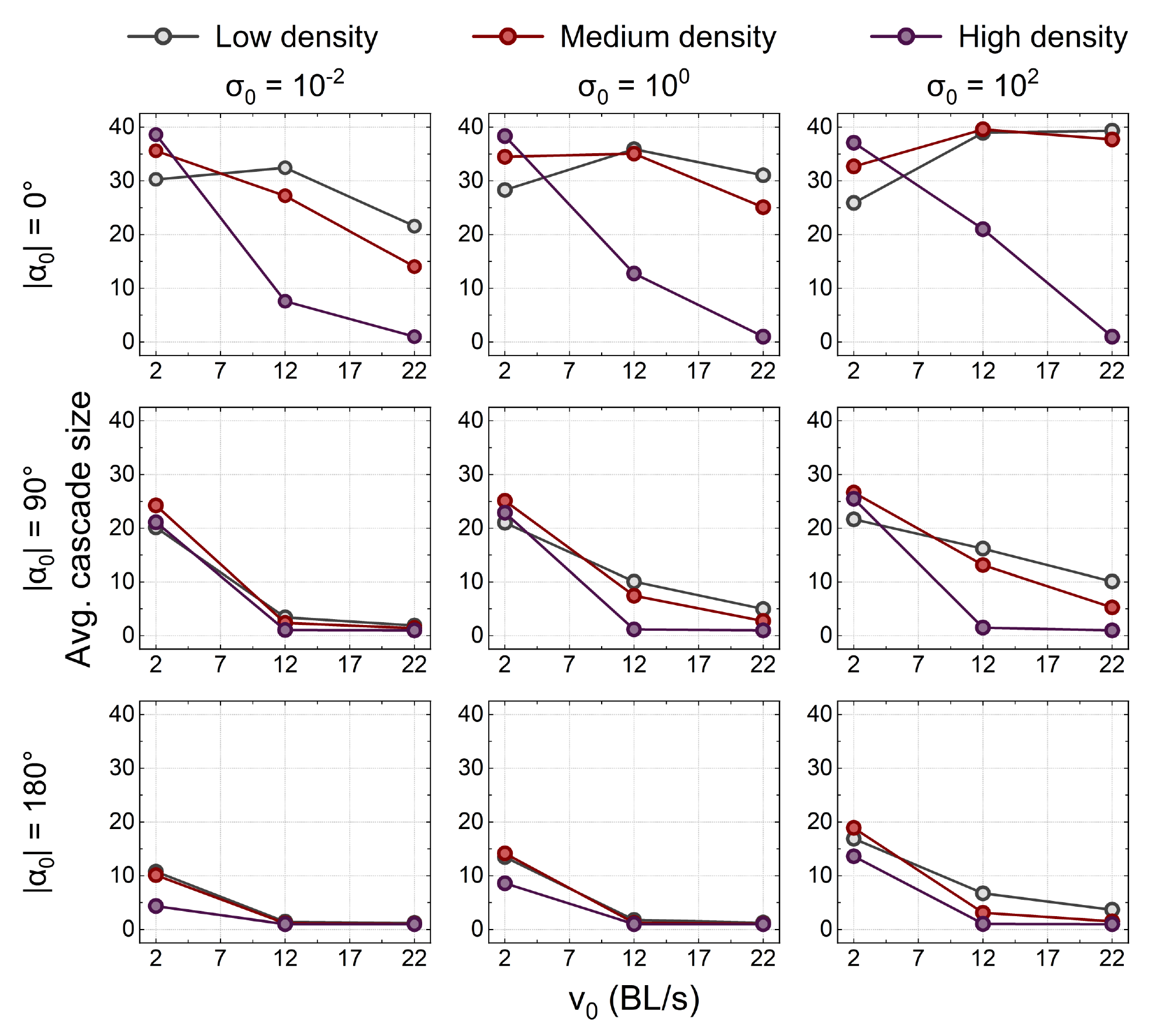}
	\begin{spacing}{1.0}
	\caption{Average cascade size as a function of constant speed $v_0$ at low, medium, and high density levels. For different noise intensities (from left to right), $\sigma_0 = 10^{-2}$, $10^{0}$, and $10^{2}$, and for different initial movement directions (from top to bottom), $|\alpha_0| = 0^\circ$, $90^\circ$, and $180^\circ$.}
	\label{Fig.4}
	\end{spacing}
\end{figure}

\begin{figure}[tbp]
	\centering
	\includegraphics[width=15cm]{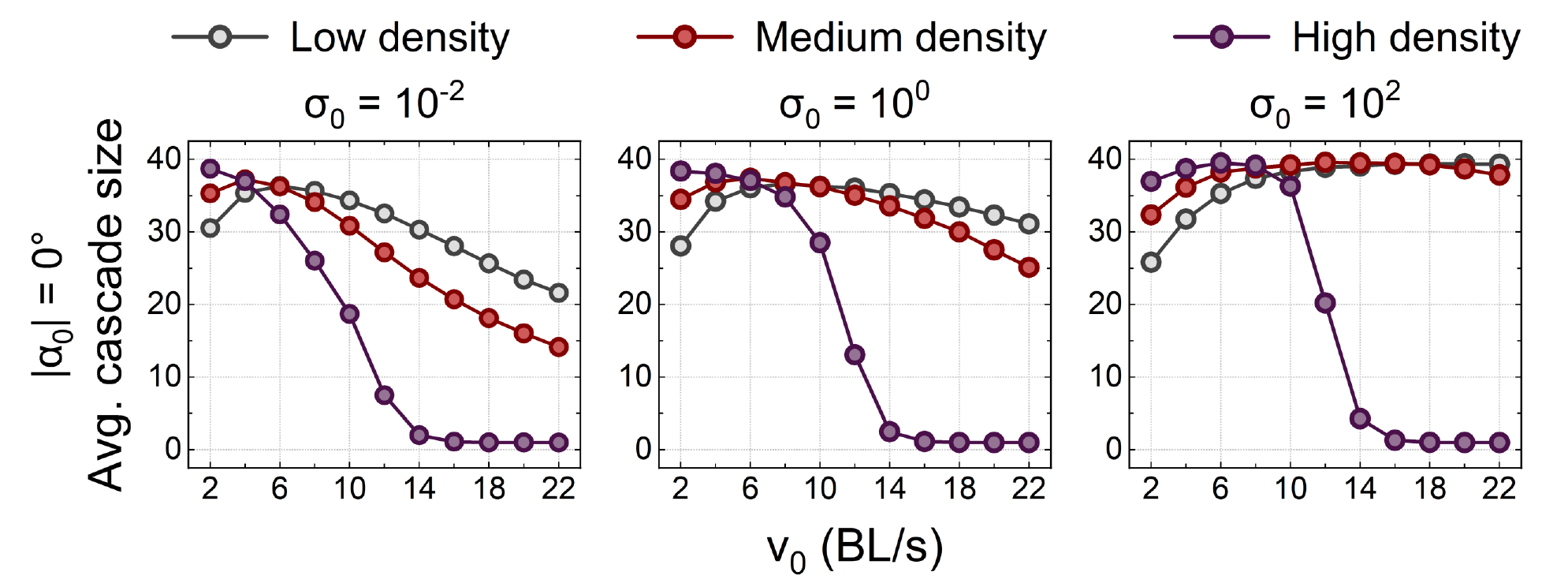}
	\begin{spacing}{1.0}
	\caption{Average cascade size as a function of constant speed $v_0$ at low, medium, and high density levels. For different noise intensities (from left to right), $\sigma_0 = 10^{-2}$, $10^{0}$, and $10^{2}$, and for the initial movement direction towards the group's center of mass, $|\alpha_0| = 0^\circ$.}
	\label{Fig.5}
	\end{spacing}
\end{figure}

\begin{figure}[tbp]
	\centering
	\includegraphics[width=15cm]{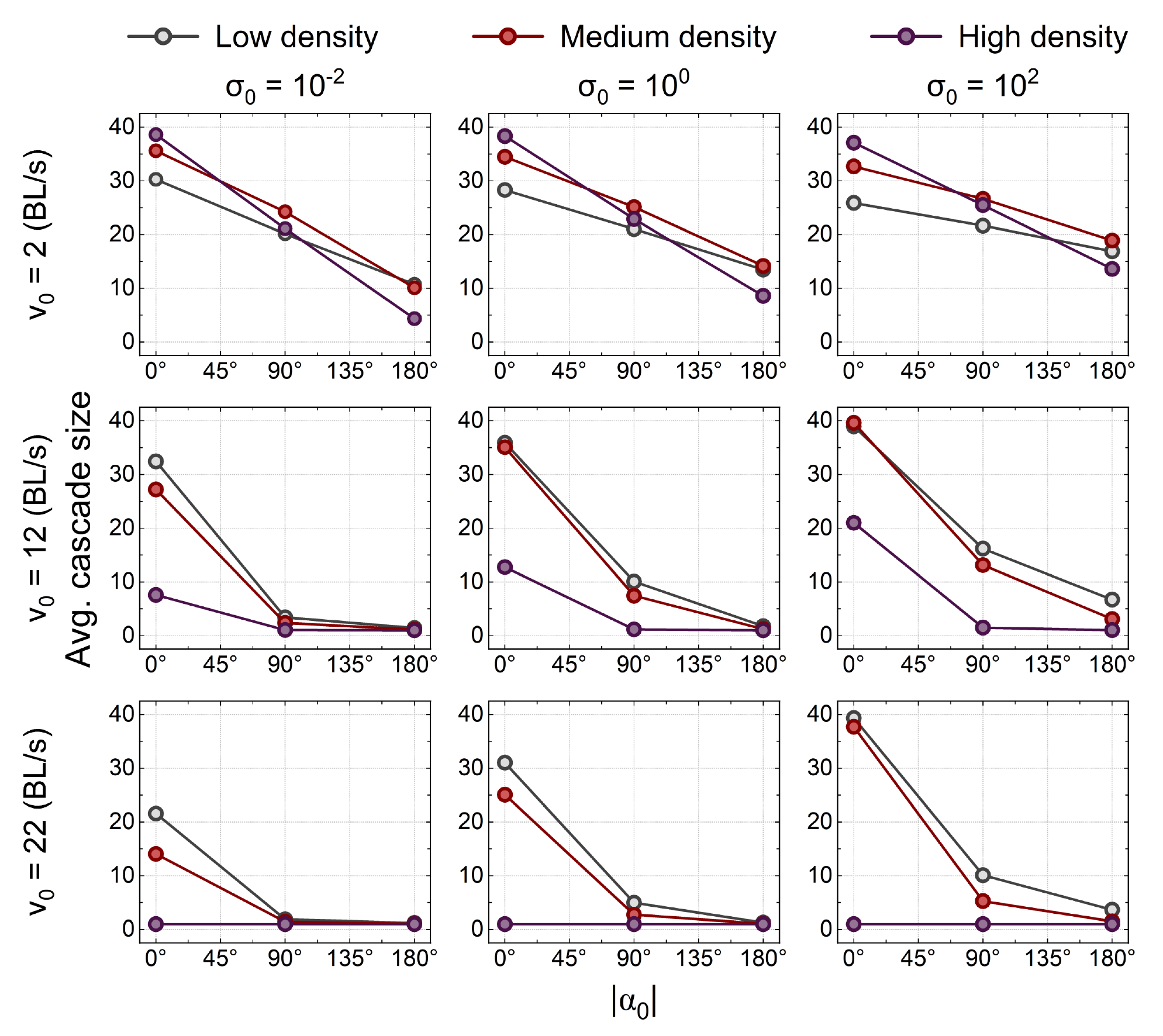}
	\begin{spacing}{1.0}
	\caption{Average cascade size as a function of initial movement direction $\alpha_0$ at low, medium, and high density levels. For different noise intensities (from left to right), $\sigma_0 = 10^{-2}$, $10^{0}$, and $10^{2}$, and for different constant speeds (from top to bottom), $v_0 = 2{\kern 1pt}BL/s$, $12{\kern 1pt}BL/s$, and $22{\kern 1pt}BL/s$.}
	\label{Fig.6}
	\end{spacing}
\end{figure}

By setting activation thresholds for the three different density levels, we investigate the effects of constant speed $v_0$, initial movement direction $\alpha_0$, and noise intensity $\sigma_0$ on average cascade size in Figs. \ref{Fig.4}-\ref{Fig.6}. Here, constant speed $v_0$ is set as $2{\kern 1pt}BL/s$ (slow speed), $12{\kern 1pt}BL/s$ (medium speed), $22{\kern 1pt}BL/s$ (fast speed); initial movement direction $\alpha_0$  is assigned as $0^ \circ$ (moving directly towards the group's center of mass), $\pm 90^ \circ$ (perpendicular to the vector towards the group's center of mass), $180^ \circ$ (moving directly away from the group's center of mass); and noise intensity $\sigma_0$  corresponds to $10^{-2}$ (low noise), $10^{-0}$ (medium noise), $10^{2}$ (high noise).

In Fig. \ref{Fig.4}, the average cascade size as a function of constant speed $v_0$ at low, medium, and high density levels is shown (see also Supplementary Videos 1 and 2). Note that when the initially startled individual moves towards the group's center of mass ($|\alpha_0| = 0^\circ$), the propagation of behavioral cascades varies significantly across different density levels and noise intensities, therefore we performed here a more detailed analysis with more sampling points with respect to the constant speed as a control parameter (see Fig. \ref{Fig.5}). For low and medium density levels, and for low and medium noise intensities ($\sigma_0 = 10^{-2}$ and $\sigma_0 = 10^{0}$), the average cascade size has its peak value at slow speeds, but for high noise intensities ($\sigma_0 = 10^{2}$) the average cascade size increases with startling speed, and reaches saturation (mostly global cascades, Fig. \ref{Fig.4} top row \& Fig. \ref{Fig.5}). For high density levels, the average cascade size decreases as a sigmoid function with the constant speed regardless of changes in noise intensity (Fig. \ref{Fig.5}). Nonetheless, when the initially startled individual moves in a direction other than towards the group's center of mass ($|\alpha_0| = 90^\circ$ and $|\alpha_0| = 180^\circ$), the average cascade size appears to decrease monotonously with increasing speed (Fig. \ref{Fig.4} middle and bottom row). Lower densities weaken the sensitivity of the cascade size to speed, because the dispersed spatial distribution limits the number of susceptible neighbors for close contacts. Besides, higher noise intensities promote the spreading of behavioral cascades, since random movement directions increase the probability that activated individuals approach susceptible neighbors. In summary, these results suggest that the increasing of the constant speed in most cases has an inhibitory effect on the propagation of behavioral cascades.

Turning now to the impact of initial movement direction $\alpha_0$ on behavioral propagation in Fig. \ref{Fig.6} (see also Supplementary Video 3). Overall, the average cascade size decreases significantly as the initially startled individual changes from moving towards the group's center of mass ($|\alpha_0| = 0^\circ$) to moving away from it ($|\alpha_0| = 180^\circ$). For the case of slow constant speed ($v_0 = 2{\kern 1pt}BL/s$), lower densities have a weakening impact on the relationship between average cascade size and initial movement direction (lower slope of the lines in Fig. \ref{Fig.6}, top row). This is because activated individuals moving at slow speeds have difficulty in contacting more distant neighbors due to the large inter-individual distance at low density levels. We also note that a similar effect exists for higher noise intensities, as the increasing directional stochasticity, results in a diffusive spread of activation, which makes behavioral contagion less sensitive to (initial) directionality of the escape movement. In contrast, for faster speeds ($v_0 = 12{\kern 1pt}BL/s$ and $v_0 = 22{\kern 1pt}BL/s$), lower densities instead increase the average cascade size for the same movement direction of the initially startled individual (e.g., $|\alpha_0| = 0^\circ$, Fig. \ref{Fig.6}). In this case, activated individuals moving at fast speeds will not leave the group quickly, which further increases the probability of cascade propagation. If the noise intensity is relatively high, activated individuals have a wider range of close contacts with susceptible neighbors in random directions, and this makes the growth in average cascade size more pronounced. The above findings demonstrate that the cascade size depends strongly on the movement direction of the initially startled individual.

\begin{figure}[tbp]
	\centering
	\includegraphics[width=15cm]{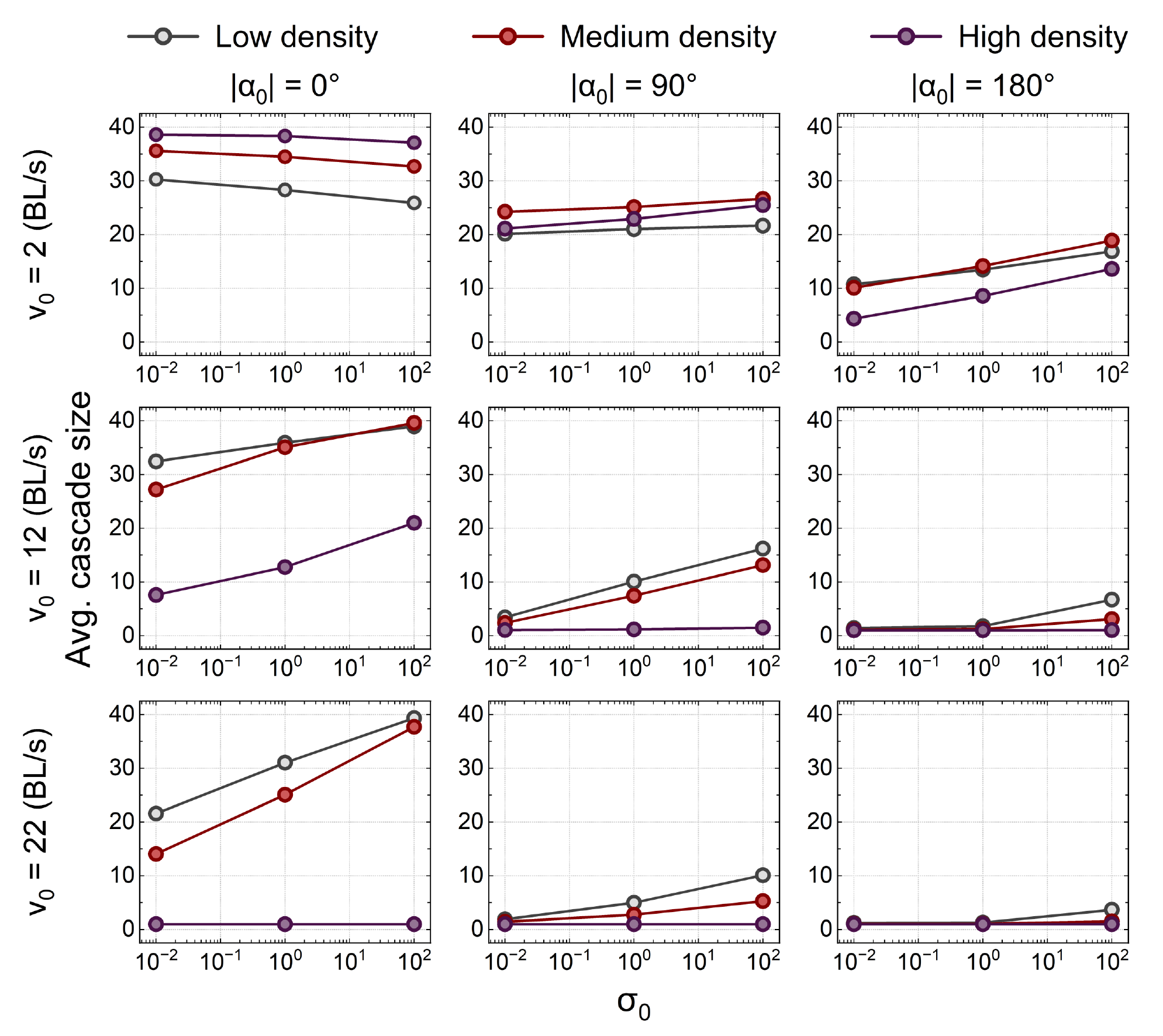}
	\begin{spacing}{1.0}
	\caption{Average cascade size as a function of noise intensity $\sigma_0$ at low, medium, and high density levels. For different initial movement directions (from left to right), $|\alpha_0| = 0^\circ$, $90^\circ$, and $180^\circ$, and for different constant speeds (from top to bottom), $v_0 = 2{\kern 1pt}BL/s$, $12{\kern 1pt}BL/s$, and $22{\kern 1pt}BL/s$.}
	\label{Fig.7}
	\end{spacing}
\end{figure}

Last, as shown in Fig. \ref{Fig.7}, we demonstrate how different noise intensities $\sigma_0$ affect the propagation of behavioral cascades (see also Supplementary Video 4). If the constant speed is relatively slow ($v_0 = 2{\kern 1pt}BL/s$), the effect of noise intensity on the average cascade size is related to the movement direction of the initially startled individual. As the initially startled individual changes from moving towards the group's center of mass ($|\alpha_0| = 0^\circ$) to moving away from it ($|\alpha_0| = 180^\circ$), higher noise intensities gradually shift from inhibiting the behavioral contagion to facilitating it. The fact is that more randomness prevents effective propagation in the movement direction of the initially startled individual. As a result of this diffusion effect, the average cascade size can be weakened if the active individual moves towards the group's center of mass, but enhanced for the case away from it. However, if the speed becomes faster ($v_0 = 12{\kern 1pt}BL/s$ and $v_0 = 22{\kern 1pt}BL/s$), independent on the movement direction of the initially startled individual, we observe larger average cascade sizes. This kind of facilitation becomes more significant with decreasing density level, since activated individuals with faster speeds will quickly leave the spatial range of the group if the density is higher, which causes the effect of directional randomness to become weaker. From this, it can be concluded that large variability in the direction of individual escape movements (rotational noise), i.e., low escape direction alignment, will typically promote the spread of behavioral contagion through spatial groups.

\section*{Conclusion} 
In this paper, we propose a spatially-explicit, agent-based model for the coupling of the behavioral contagion process to the (adaptive) interaction network dynamics to study collective evasion behavior in fish groups. In particular, in the vicinity of the critical response threshold for the phase transition of the system from local to global cascades, we find that movement parameters (startle speed, initial directionality, and directional noise) play a crucial role in regulating the propagation of behavioral cascades. Specifically, we reveal that the range of behavioral contagion is suppressed by increasing constant speed in most cases, but may be promoted or peaked at different density levels and noise intensities if the initially startled individual moves towards the group's center of mass. The cascade size depends strongly on the movement direction of the initially startled individual, which decreases from moving towards the group's center of mass to moving away from it. Slow speeds, lower densities, and higher noise intensities make behavioral contagion less sensitive to directionality, while instead promoting larger cascades with increasing speed. Large variability in the direction of individual escape movements (rotational noise) typically promotes the spread of behavioral cascades through spatial groups, and its impact becomes more pronounced as the density decreases. However, higher noise intensities lead to a reduction in cascade size at slow speeds if the initially startled individual moves towards the group's center of mass. These conclusions demonstrate that movement dynamics have a complex impact on collective evasion behavior.

This work advances our understanding of rapid coordinated responses and collective information processing in mobile groups, and provides valuable clues to further control and management of collective behavior. In emergency situations (e.g., fires \cite{Nilsson2009}, earthquakes \cite{Feng2020}, and terrorist attacks \cite{Shields2009}), the collective motion of large-scale crowds may suddenly change during the evacuation process. For example, if a pedestrian immediately changes direction after finding a safer escape route or perceives the danger, his or her surrounding neighbors in a panic will have a high probability of imitating this behavior \cite{Hasan2011}. However, the sudden change in behavior may disrupt the order of collective motion, give rise to more physical collisions, and even cause serious stampedes. Therefore, the potentially disastrous spread of behavioral cascades can be reduced by designing effective information transmission and guidance strategies to improve the efficiency and safety of crowd evacuation. To adapt to dynamic changes in surrounding environments and achieve coordinated maneuvers within swarms of robots when handling complex tasks \cite{Dorigo2020,Garattoni2018}, continuous adjustment of movement directions is required to support path optimization and decision making \cite{Min2011}. Inspired by the principle of collective evasion behavior in fish groups, we can use dynamic parameters to control the movement direction in swarming robots. For example, when the individual closest to an obstacle perceives the positional information, we can use a control algorithm to adjust its movement direction and constant speed, and spread this behavior to a part or the whole group for obstacle avoidance. This kind of bionic swarm robotics approach has the advantages of high adaptability, efficiency, and flexibility, and can better deal with the requirements of challenging tasks in complex environments.

Overall, this work explores the dynamic mechanisms of collective escape behavior in mobile animal groups, reveals the important influence of movement parameters on behavioral propagation, and provides new perspectives into our understanding of collective behavior. In future work, we plan to further improve the motion equations (e.g., force-based \cite{Zhao2022}) of individuals to describe the interactions with other individuals and the environment more accurately, which allows precise simulation and prediction for the dynamic evolution of behavioral cascades. It is also worth exploring the effect of other kinetic parameters on behavioral contagion and how these parameters can be adjusted to optimize group coordination and adaptability to help effectively organize and control collective evasion behavior. Importantly, the visual network needs to be recalculated at each time step due to movement behavior, which results in high computational costs. Hence, subsequent work could try to explore more efficient algorithms or optimize the calculation process to improve the real-time performance of this model. We expect that our work will inspire more general models of behavioral contagion and pave the way for deeper insights into rapid coordinated collective responses in biological and social systems.

\section*{Data availability} 
The data supporting the results within this paper are available from the corresponding author upon reasonable request.

\section*{Code availability} 
The code used for the crowd evacuation simulations within this paper is available from the corresponding author upon reasonable request.

\bibliographystyle{naturemag}	
\bibliography{references}

\section*{Acknowledgements} 
WW and XZ acknowledge funding by the National Major Scientific Research Instrument Development Project under Grant 61927804. PR acknowledges funding by the Deutsche Forschungsgemeinschaft (DFG, German Research Foundation) under Germany’s Excellence Strategy – EXC 2002/1 “Science of Intelligence” – project number 390523135.

\section*{Author contributions} 
PR and WW conceived and designed the research. PR \& WW analyzed relevant data. PR \& WW planned the numerical simulations. WW wrote the numerical code and performed the simulations.  WW wrote the manuscript, and PR revised it and provided editorial input.   
XZ provided resources, funding acquisition, and project administration.

\section*{Competing interests}
The authors declare no competing interests.

\section*{Additional information}
\textbf{Correspondence} and requests for materials should be addressed to P.R.

\clearpage
\section*{Supplementary Videos}

\paragraph{Supplementary Video 1 -- Impact of speed $v_0$ at low density:} Visualization of example simulations for different escape speeds $v_0=2{\kern 1pt}BL/s$ (slow), $v_0=12{\kern 1pt}BL/s$ (medium), and $v_0=22{\kern 1pt}BL/s$ (fast); Parameters: density $\rho=0.1{\kern 1pt}BL^{-2}$, initial movement direction $\alpha_0=0^\circ$, noise intensity $\sigma_0=1$, all remaining parameters as in Tab. \ref{tab:parameters}. 

\paragraph{Supplementary Video 2 -- Impact of speed $v_0$ at high density:} Visualization of example simulations for different escape speeds $v_0=2{\kern 1pt}BL/s$ (slow), $v_0=12{\kern 1pt}BL/s$ (medium), and $v_0=22{\kern 1pt}BL/s$ (fast); Parameters: density $\rho=0.4{\kern 1pt}BL^{-2}$, initial movement direction $\alpha_0=0^\circ$, noise intensity $\sigma_0=0.2$, all remaining parameters as in Tab. \ref{tab:parameters}. 

\paragraph{Supplementary Video 3 -- Impact of initial movement direction $\alpha_0$ at medium density:} Visualization of example simulations for different initial movement directions $\alpha_0=0^\circ$ (towards center of mass), $\alpha_0=90^\circ$ (sideways), and $\alpha_0=180^\circ$ (away from center of mass); Parameters: density $\rho=0.18{\kern 1pt}BL^{-2}$, speed $v_0=2{\kern 1pt}BL/s$, noise intensity $\sigma_0=0.2$, all remaining parameters as in Tab. \ref{tab:parameters}.

\paragraph{Supplementary Video 4 -- Impact of noise intensity $\sigma_0$ at medium density:} Visualization of example simulations for different noise intensities $\sigma_0=0.2$ (low), $\sigma_0=1$ (medium), and $\sigma_0=100$ (high); Parameters: density $\rho=0.18{\kern 1pt}BL^{-2}$, speed $v_0=22{\kern 1pt}BL/s$, initial movement direction $\alpha_0=0^\circ$, all remaining parameters as in Tab. \ref{tab:parameters}

% \section*{Figures} 

\end{document}